\documentclass[sigconf]{acmart}
\AtBeginDocument{%
  }

\copyrightyear{2026} 
\acmYear{2026} 
\setcopyright{cc}
\setcctype{by}
\acmConference[BUILDSYS ’26]{The 13th ACM International Conference on Systems for Energy-Efficient Buildings, Cities, and Transportation}{June 22--25, 2026}{Banff, Canada} 
\acmBooktitle{The 13th ACM International Conference on Systems for Energy-Efficient Buildings, Cities, and Transportation (BUILDSYS '26), June 22--25, 2026, Banff, Canada}
\acmISBN{978-1-4503-XXXX-X/2018/06}

\usepackage{booktabs}
\usepackage{multirow}

\begin{document}

\title{Thermal-GEMs \\ Generalized Models for Building Thermal Dynamics}


\author{Felix Koch}
\affiliation{%
  \institution{Technical University of Applied Sciences Rosenheim}
  \city{Rosenheim}
  \country{Germany}}
\email{felix.koch@th-rosenheim.de}

\author{Fabian Raisch}
\authornote{Main affiliation at Technical University of Applied Sciences Rosenheim; doctoral candidate at Technical University of Munich (cooperative doctorate with Technical University of Applied Sciences Rosenheim).}
\affiliation{%
  \institution{Technical University of Applied Sciences Rosenheim}
  \city{Rosenheim}
  \country{Germany}}
  \affiliation{%
\institution{Technical University of Munich}
  \city{Munich}
  \country{Germany}}

\author{Benjamin Tischler}
\affiliation{%
  \institution{Technical University of Applied Sciences Rosenheim}
  \city{Rosenheim}
  \country{Germany}}

\renewcommand{\shortauthors}{Koch et al.}

\begin{abstract}

Data-driven models for building thermal dynamics are a scalable approach for enabling energy-efficient operation through fault detection \& diagnosis or advanced control. 
To obtain accurate models, measurement data from a target building spanning months to years are required. 
Transfer Learning (TL) mitigates this challenge by employing pretrained models based on single or multiple source buildings. General multi-source TL models promise to outperform single-source TL, but alternative multi-source modeling architectures remain to be explored, and evaluation on real-world data is missing. Moreover, time series foundation models (TSFM) have emerged as candidates for the best-performing general models. 
Hence, we conduct a first, comprehensive assessment of general modeling approaches for building thermal dynamics, including multi-source TL and TSFMs. 
Our assessment includes ablations using four state-of-the-art multi-source TL architectures and evaluations on synthetic as well as real-world data. We demonstrate that multi-source TL models are highly effective in accurately modeling buildings in real-world applications, yielding up to 63\% lower forecasting errors compared to single-source TL. 
Moreover, our results suggest a trade-off between multi-source TL models exclusively pretrained with building data and TSFMs pretrained with a multitude of different time series, revealing that data from 16–32 source buildings must be available over 1 year for pretraining multi-source TL models to consistently outperform TSFMs as evaluated using the mean absolute error. These findings provide practical guidance for selecting modeling strategies based on the number of source buildings available for pretraining multi-source TL models.
\end{abstract}



\received{06 February 2026}
\received[accepted]{04 April 2026}

\maketitle

%
%
\section{Introduction}
Data-Driven Models (DDMs) are widely adopted for applications in energy-efficient operations, including Model-Predictive control (MPC) \cite{drgovna2020allyouneedtoknowmpc, balali2023energyddmcontrol, lee2022aimodelwithmpc}, Reinforcement Learning (RL) \cite{arroyo2022rlmpc, CORACI2023117303}, and fault detection and diagnosis (FDD) \cite{CHEN2023121030, DU20091624}. In particular, Machine Learning (ML) models are used as DDMs because they offer scalable implementation, strong predictive performance, and no requirement for extensive domain-specific knowledge, as demonstrated in \cite{choi2023performance, elmaz2021cnnlstmforecastthermal} for thermal dynamics in buildings. However, ML-based models often require months to years of data to achieve satisfactory prediction performance \cite{raisch2025cl}. In reality, little to no data is available when advanced control or FDD systems are installed. Additionally, each building requires its own specifically fitted DDM, as buildings are inherently heterogeneous. Hence, a unique DDM is needed for each target building, with little or no data available. To address these problems, Transfer Learning (TL) approaches have emerged. TL employs a pretrained source model to model a specific target building, thereby addressing inaccurate forecasts caused by data sparsity.  \\

In the domain of thermal dynamics of buildings, a single-source TL approach is commonly used \cite{chen2020transfer, pinto2022transferleaningsmartbuildingsinglesource}. In single-source TL, a model is pretrained on a single source building and then fine-tuned on the desired target building, sharing the same number and type of sensors as the source building. Raisch et al. \cite{GenTL2025} were the first to demonstrate a multi-source TL approach. For the multi-source approach, a single DDM is trained on data from multiple sources, which we will refer to as a multi-source general model (MSGM). \cite{GenTL2025} used a long short-term memory (LSTM) as DDM; however, it is unclear whether this is the best choice. In addition, they used simulated data to demonstrate that an MSGM outperforms single-source TL for thermal dynamics in buildings. To build trust in a practical deployment in real buildings, a comparison between the MSGM approach and single-source TL on real-world data is needed \cite{tian2018reviewrealworldbuildings, batty2018digitaltwins}.  \\

\noindent\textbf{Research Question 1:} How does multi-source TL perform compared to a single-source approach for building thermal dynamics on real-world data, and which architecture is best suited for an MSGM? \\

Recently, Time Series Foundation Models (TSFM) have emerged as an alternative to MSGMs as general models \cite{mulayim2024tsfmbuildingsrevolutionize,liang2024enablingtsfmcontrastivecurr}. TSFMs are pretrained on vast amounts of heterogeneous data from different domains to learn efficient representations of temporal patterns. TSFMs can be advantageous in domains with limited data or varying available features (i.e., different types and numbers of sensors). For thermal dynamics in buildings, this was partially explored by \cite{mulayim2024tsfmbuildingsrevolutionize}. However, two aspects remain unexplored: First, the use of state-of-the-art multivariate TSFM applied to thermal dynamics. Second, and even more importantly, TSFMs have not yet been compared to MSGMs for building thermal dynamics, leaving the question of their effectiveness open. \\

\noindent \textbf{Research Question 2:} How do current state-of-the-art TSFMs perform for building thermal dynamics compared to an MSGM? \\

In practice, buildings have different types and numbers of sensors and measurement rates. Currently, training an MSGM requires data from multiple buildings with the same type and number of sensors, with the same measurement rate. For the scope of this paper, we call these buildings \textit{same sensor buildings}.
Because there are many different sensor configurations, it becomes difficult to gather enough same sensor buildings for each sensor combination to train an MSGM. We refer to this challenge as the \textit{same sensor building sparsity}. This issue is not present for TSFMs, which benefit from large, heterogeneous training corpora across many domains and can handle various feature input and output formats. Assuming that using more source buildings improves multi-source DDM performance leads to the following unexplored question: \\

\noindent \textbf{Research Question 3:} How many source buildings are required for pretraining for a multi-source general model to outperform current TSFMs? \\

By addressing these research questions, we hope to enhance energy-efficient building operations by advancing general models for building thermal dynamics tailored to the frequent scenario of data sparsity, i.e., little or no real measurement data are available for a target building.


\begin{figure}[htbp]
  \centering
  \includegraphics[width=\linewidth]{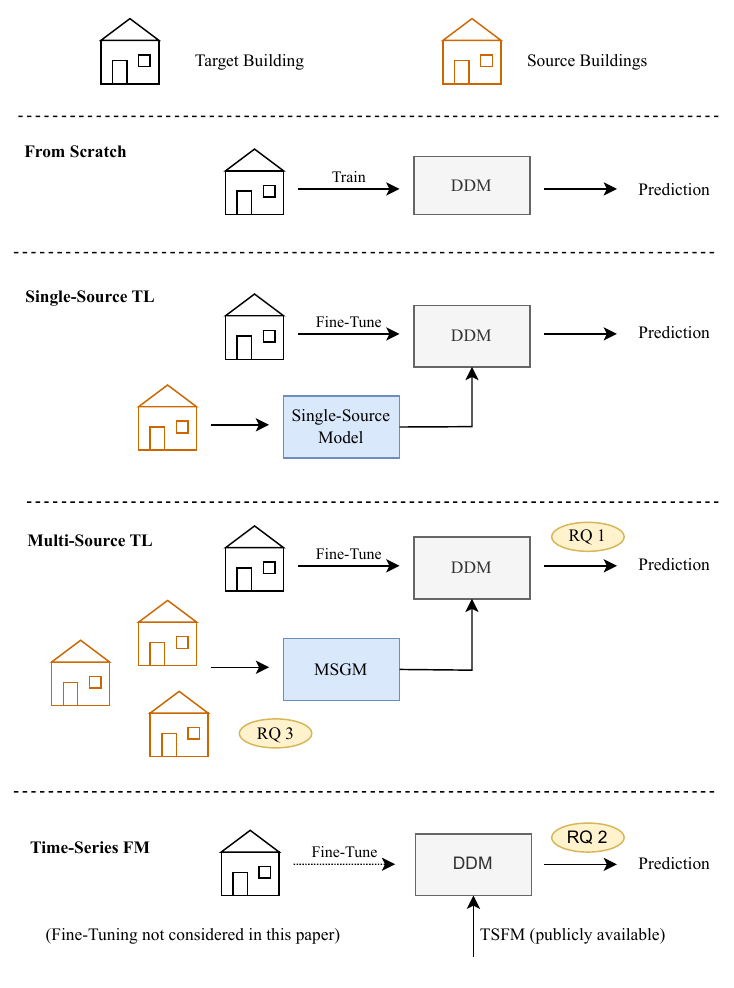}
  \caption{Overview of modeling strategies regarding our research questions.}
  \label{fig:rq_overview}
\end{figure}

\section{Related Work}
\label{sec:related_work}
Addressing data sparsity for buildings with Transfer Learning (TL) has received significant attention in recent years.
Chaudhary et al. \cite{chaudhary2025transfer}, for example, evaluated a single-source TL approach for modeling Norwegian buildings and found significant performance improvements compared to training a dedicated model from scratch. 
Also, Duo et al. \cite{dou2025transfer} adopted a similar approach. They modeled thermal comfort and electric load of buildings with single-source TL from synthetic source data to a real-world target. They also found significant improvements for the TL approach over modeling from scratch. This improvement was found to be larger with more prevalent data sparsity. However, the study used only one target building and evaluated only 3 months of the summer season.
Pinto et al. \cite{pinto2022sharingiscaring} evaluated single-source strategies on over 250 simulated buildings. The authors showed that TL yields performance increases when the source and target buildings share the same climatic conditions, even when the occupancy schedules differ and under low data availability circumstances. However, their results are only based on simulated data and do not include tests on real-world buildings. 
Li et al. \cite{li2024building} conducted experiments on deep learning architectures for single-source TL. Complementary to \cite{pinto2022sharingiscaring}, the authors verified that the effectiveness of single-source TL can be transferred to real-world buildings.
Raisch et al. \cite{GenTL2025} used data from 450 synthetically generated central European buildings to train a multi-source LSTM model. Their experiments showed that an MSGM outperforms both single-source TL models and models trained from scratch for building thermal predictions. The authors reported their results using seasonal fine-tuning, recognizing seasonal differences in building dynamics. With this approach, the results of \cite{GenTL2025} are robust to seasonal differences, effects that are mostly overlooked in the aforementioned literature.  \\
All studies comparing multi-source modeling approaches with single-source strategies conclude that using multiple buildings for training is advantageous. However, it remains unclear to what extent this can be transferred into a real-world thermal prediction scenario. Although \cite{li2024building} showed that this transfer can be done for the single-source TL approach, this has not yet been done for multi-source TL as requested by \cite{hertel2023transformer, GenTL2025}. \\

Other approaches to tackle data sparsity in buildings relied on Time Series Foundation Models instead.\\
Liang et al. \cite{liang2024enablingtsfmcontrastivecurr}, for instance, tested the two TSFMs Tiny Time Mixer \cite{ekambaram2024tinytimemixer} and Chronos \cite{ansari2024chronos1} on three distinct building energy datasets (electric load forecasting). The authors observed that in a few-shot setting, both TSFMs outperformed state-of-the-art deep learning models from scratch, like the LSTM \cite{hochreiter1997lstm}, Temporal Fusion Transformer \cite{lim2021temporalfusiontransformertft}, and Autoformer \cite{wu2021autoformer}. The authors also emphasize that fine-tuning TSFMs is hard, motivating the use of either zero-shot forecasting or contrastive curriculum learning. However, \cite{liang2024enablingtsfmcontrastivecurr} did not account for thermal dynamics in buildings and also did not consider an MSGM as a baseline. 
Park et al. \cite{park2025probabilistictsfmenergy} used the TSFMs Chronos \cite{ansari2024chronos1}, TimesFM \cite{das2024timesfm}, and Moirai \cite{woo2024moirai} for a building energy forecasting task on a net-zero commercial building against state-of-the-art deep learning approaches (such as the Temporal Fusion Transformer \cite{lim2021temporalfusiontransformertft}). The authors found that with Low-Rank Adaptation (LoRA) fine-tuning, the TSFMs outperform the deep learning forecasting models across multiple building zones and seasonalities.
Additionally, Mulayim et al. \cite{mulayim2024tsfmbuildingsrevolutionize} evaluated several state-of-the-art TSFMs for modeling building thermal dynamics and electric forecasting on real-world datasets. The authors found that TSFMs outperform AutoARIMA \cite{seabold2010autoarima} in multiple cases. Still, they only compared TSFMs to each other and did not account for MSGM approaches for the same tasks. Also, \cite{mulayim2024tsfmbuildingsrevolutionize} focused on univariate forecasting in their work. \\
Current related work consistently indicates that TSFMs are promising for modeling buildings under sparse data conditions. However, it remains unclear to what extent these findings generalize to the modeling of thermal dynamics in buildings. Moreover, most of the aforementioned studies focus exclusively on univariate TSFMs.
\footnote{
Here, we use multivariate as an umbrella term when models incorporate variables other than the target variable as inputs. 
\cite{mulayim2024tsfmbuildingsrevolutionize} use LagLlama \cite{rasul2024lagllama} and TimesFM \cite{das2024timesfm} and claim, to our understanding erroneously, that these models are multivariate.
As \cite{mulayim2024tsfmbuildingsrevolutionize} provides limited detail on covariate integration, we assume that their univariate forecasts did not include covariate support. 
Furthermore, their evaluation of MOMENT \cite{goswami2024moment} in a zero-shot setting is problematic, as the model's randomly initialized forecasting heads require fine-tuning to produce non-random results. Thus, we did not include MOMENT in our evaluations, although it achieved promising results on the Ecobee dataset \cite{luo2022ecobee} in \cite{mulayim2024tsfmbuildingsrevolutionize}.
}
Multivariate TSFMs may exhibit fundamentally different behavior and therefore require separate investigation. The relative performance of TSFMs compared to MSGMs remains an open question.

The existing literature lacks comprehensive evaluations of MSGMs for building thermal dynamics on real-world datasets. Moreover, comparisons between TSFMs and MSGMs techniques for this task remain unexplored. Additionally, the role of same-sensor building sparsity and its influence on the performance of MSGMs have not yet been investigated. 

To address this research gap, we compare state-of-the-art deep learning models for the MSGMs with three state-of-the-art TSFMs. For the MSGM, we investigate four widely used deep learning architectures. To ensure comprehensive assessments and robust results, these models are trained and evaluated on three distinct building datasets, which together represent more than 500 homes in the US, Canada, the UK, Germany, France, and more. To further strengthen the robustness of the analysis, we also account for seasonal differences in building dynamics. Moreover, we improve the robustness of our experiments compared to the existing literature by using a more representative seasonal fine-tuning split. 

The authors see the following contributions within the paper:

\begin{enumerate}
    \item We present the first investigation of multi-source general models for building thermal dynamics using real-world data.
    \item We provide the first comprehensive comparison between two types of general models, state-of-the-art TSFM models and MSGMs for building thermal dynamics.
    \item We perform the first analysis on the number of buildings required for pretraining a multi-source general model to outperform TSFMs.
\end{enumerate}


The remainder of this paper is structured as follows: In the following Section \ref{sec:method}, we explain the method of our paper. This is followed by the experiments in Section \ref{sec:experiments}. Last, we discuss our results in Section \ref{sec:discussion} and draw a conclusion in Section \ref{sec:conclusion}. We share the models and code to reproduce all results on GitHub [anonymous].

\section{Method}
\label{sec:method}
In this section, we describe the method of our paper. We first introduce the data sets for our study in Section \ref{sec:data}. Thereafter, in Section \ref{sec:gm_training}, we explain the training process of the multi-source general models. This is followed by a description of our seasonal finetuning in section \ref{sec:seasonal_fineuning}. Thereafter, we introduce TSFMs in Section \ref{sec:tsfms}. Lastly, we explain how we assess seasonality in Section \ref{sec:seasonal_fineuning} and the evaluation of our methods in Section \ref{sec:evaluation}.

\begin{table}[h!]
\centering
\caption{Features included in the three datasets.}
\label{tab:dataset_features}

\begin{tabular}{l| l | c | c | c}
\toprule
\multicolumn{2}{c|}{\textbf{Observable}} & \multicolumn{3}{c}{\textbf{Dataset}} \\
\hline
\textbf{Name} & \textbf{Type} & \textbf{D1} & \textbf{D2} & \textbf{D3} \\
\hline
Indoor Humidity & Feature & \checkmark & \(\times\) & \(\times\) \\
Thermostat Heat Setpoint & Feature & \checkmark & \(\times\) & \checkmark \\
Thermostat Cool Setpoint & Feature & \checkmark & \(\times\) & \(\times\)\\
Supply Air Fan Runtime & Feature & \checkmark & \(\times\) & \(\times\) \\
Presence of Occupants & Feature & \checkmark & \(\times\) & \(\times\) \\
Heating Power & Feature & \checkmark & \checkmark & \checkmark \\
\hline
Outdoor Wind Speed & External Feature & \(\times\) & \checkmark & \checkmark \\
Outdoor Humidity & External Feature & \(\times\) & \checkmark & \(\times\) \\
Outdoor Temperature & External Feature & \checkmark & \checkmark & \checkmark \\
Solar Radiation & External Feature & \checkmark & \checkmark & \checkmark \\
\hline
Indoor Room Temperature & Target & \checkmark & \checkmark & \checkmark \\

\bottomrule

\end{tabular}
\end{table}

\subsection{Data}
\label{sec:data}

In this paper, we investigate indoor temperature forecasts in buildings. For this purpose, we use three datasets: two based on real-world sensor data and one on simulated data. In the following, we briefly explain each data set: \\

\noindent
\textbf{D1: EcoBee Dataset.} The Ecobee dataset \cite{luo2022ecobee} is a real-world public dataset from the EcoBee Donate Your Data (DYD) program. In the DYD program, occupants self-reported metadata describing their home and occupancy behavior and share their data from designated thermostats. Data were tracked from homes in four different climate zones in North America, namely Texas (TX), California (CA), New York (NY), Illinois (IL), and Ottawa (OT) in Canada. The dataset contains only single-family homes, where each time series represents an individual home. 
Additionally, we extended the dataset with solar irradiation data from nearby weather stations \cite{CopernicusC3S2017}, as this feature was missing. \\

\noindent
\textbf{D2: Ideal Households.} The Ideal household energy dataset \cite{idealdataset} comprises data from 32 homes in the UK, with a total of 255 rooms. In this case, multiple time series may originate from one home. Originally, the data source included more homes, but we filtered them out beforehand because some time series lacked heating power. We excluded parts of the time series that lacked data due to monitoring issues. The dataset includes houses, bungalows, and flats. The homes were located in Edinburgh, Mid Lothian, West Lothian, East Lothian, or Fife. As the authors of this data set did not provide external weather data, these variables were added for each building using the nearest Weather Underground data \cite{wunderground2023}, as proposed by \cite{idealdataset}.
All rooms and homes were recorded separately, resulting in different start and end times. Each home may have multiple thermal zones, with a separate time series for each zone. 
\\

\noindent
\textbf{D3: European Households.} The European Households dataset is a non-public, simulated dataset aimed at covering single-family homes present in Central Europe. The authors of \cite{raisch2025cl, GenTL2025} made this information available for this research. The dataset consists of 450 distinct buildings with different building properties, occupancy profiles, and weather data. It spans 90 variations in envelope parameters, representing homes constructed between 1949 and the present, and includes weather data from Prague, Zurich, Berlin, Belgrade, and Paris. The data set was simulated using the BuilDa simulation framework \cite{krug2025builda2, krug2025builda}.
For further details regarding the dataset, we refer to \cite{GenTL2025, raisch2025cl}. \\



\noindent
\textbf{Data Split.} We split each dataset into a source and a target set. The source set contains data used for training of the single-source and multi-source pretrained models. The target set is used to evaluate the methods.

\begin{table}[h!]
\centering
\caption{Source and target split for all datasets}
\label{tab:train_test_splits}
\begin{tabular}{l|c|c | c}
 & \textbf{EcoBee} & \textbf{Ideal} & \textbf{Euro. House} \\
 \hline
 \textbf{\# Rooms Total} & 932 & 227 & 450 \\
 \textbf{\# Sources Total} & 892 & 202 & 400 \\
 \textbf{\# Targets} & 40 & 25 & 50 \\
 \hline
 \textbf{\% Targets} & 4.3 & 11.0 & 11.1
\end{tabular}
\end{table}

Table \ref{tab:train_test_splits} lists the number of source and target time series available for each dataset. We selected a subset of target time series from each dataset so that the proportion of the corresponding buildings from each location in the subset reflects their distribution in the full dataset. The Ideal dataset is an exception to that, as it includes multiple time series for a single home (one time series per room). To avoid data leakage, we split the dataset by building, resulting in 4 buildings with a total of 25 time series in the target set. For the Ecobee data set, we selected less than 10\% time series in the target set to lower computational effort. Because the datasets had different sampling rates, all data were downsampled to 15-minute intervals.

\subsection{General Model Training}
\label{sec:gm_training}

We use the source data sets from Section ~\ref{sec:data} to train multi-source general models (MSGM). An MSGM is a data-driven model based on a deep learning architecture trained not just for one building, but for multiple buildings. Therefore, the time series for a single building is sliced into training examples. One training example consists of a lookback window (input) of length 96 (corresponding to 1 day) and the ground truth horizon-step temperature trajectory (forecast) of 4 steps (1 hour). We chose the lookback and forecast horizon based on related literature \cite{GenTL2025, pinto2022sharingiscaring,chen2020transfer,li2024building}. After slicing, we shuffle the training example across all source buildings in the dataset. We chose AdamW \cite{loshchilov2017adamw} as the optimizer because we observed better generalization behavior than Adam \cite{kingma2014adam}.
We reduce the number of time series for training the MSGMs to 200 to ensure comparability and computational feasibility. From those 200 time series, we select at random 180 for the training set and 20 for the validation set. 

We selected four deep learning architectures that are frequently used in the literature and have demonstrated good performance on similar tasks as potential general models.  \\

\noindent \textbf{A1: Long Short-Term Memory (LSTM).} The LSTM \cite{hochreiter1997lstm} is a recurrent neural network architecture with additional gates to retain information over long sequence lengths. They are already widely used to model thermal dynamics in buildings, as described in \cite{choi2023performance,raisch2025cl}. In \cite{GenTL2025}, LSTMs were already used for an MSGM. 
\\

\noindent \textbf{A2: Transformer.} The Transformer is a widely used architecture that relies on a self-attention mechanism to capture global dependencies. 
In \cite{choi2023performance}, the transformer was found most promising for their case study on thermal dynamics in buildings.
We chose to use a standard Transformer as proposed in \cite{vaswani2017attentionisallyouneed} instead of other Transformer-based architectures, such as Informer \cite{zhou2021informer} or PatchTST \cite{Yuqietal-2023-PatchTST}, based on the results of \cite{hertel2023transformer}. We added a cosine positional embedding to the features, as described in \cite{vaswani2017attentionisallyouneed}, to mitigate potential encoder invariance. \\

\noindent \textbf{A3: Mamba.} Mamba \cite{gu2023mamba} is a structured state-space model that captures long-term dependencies by retaining important information with its selection mechanism. Mamba has shown comparable performance to attention-based methods in time series forecasting \cite{wang2025mambatimeseriesforecasting}, while requiring less computational cost than Transformers due to its sublinear complexity with respect to input sequence length, enabling significantly larger context windows. 
Therefore, it is a reasonable choice to explore this architecture for thermal dynamics in buildings. \\

\noindent\textbf{A4: xLSTM.} xLSTM \cite{beck2024xlstm} is an extension to the original LSTM \cite{hochreiter1997lstm}, introducing exponential gating and improved memory strategies with sLSTM and mLSTM layers. As LSTMs are already widely used to model thermal dynamics, it is reasonable to include their extension, xLSTM, in this study. We used an sLSTM as the first layer and filled the rest with mLSTM layers. The rationale behind it is that the sLSTM captures long-term dependencies, while the mLSTM enables the modeling of more complex patterns. \\
 
For each architecture, we use Bayesian Optimization with Optuna \cite{akiba2019optuna} (see Appendix Table \ref{tab:z_hyperparam_search_gms} and \ref{tab:gm_hyperparams}) to tune the hyperparameters. 
For all experiments, we selected the model we used for further experiments based on the lowest RMSE on the validation set. Because differences can arise from random initialization, we trained each general model with the best hyperparameters on each dataset 4 times.

\subsection{Seasonal Fine-Tuning}
\label{sec:seasonal_fineuning}

\begin{figure}[]
  \centering
  \includegraphics[width=\linewidth]{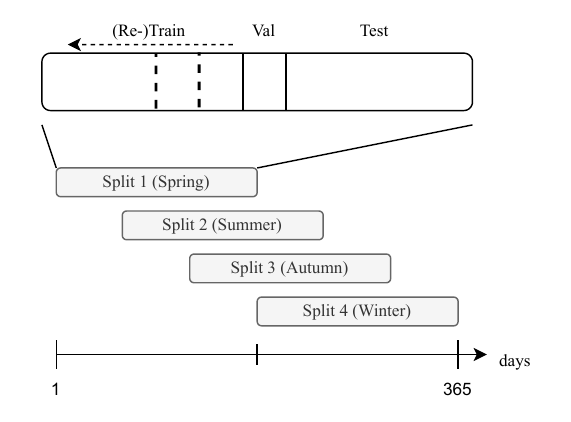}
  \caption{Seasonal Fine-Tuning split for the experiments}
  \label{fig:ft_splits}
\end{figure}

After training an MSGM, it can be fine-tuned/applied to an unseen target building. Usually, little to no data is used for this process. However, performance may vary significantly depending on the season in which the model is applied \cite{choi2023performance}. These differences can be explained by factors such as outdoor temperature and solar irradiance, which vary with seasonality.
To counteract biased fine-tuning results due to seasonal effects, we evaluate the models using a seasonal fine-tuning split inspired by \cite{GenTL2025}, as illustrated in Figure \ref{fig:ft_splits}. 
The 1-year time series is divided into four seasonal segments. Each segment contains consecutive train, validation, and test sets. The training set can vary to reflect different data sparsity scenarios (ranging from 0.5 to 2.5 months). The validation set always spans 0.5 months, followed by a 3-month test set.

The model is fine-tuned on the corresponding training set. For fine-tuning, we initialize the weights with the pretrained model. This process, called weight initialization, was found to be better than layer freezing \cite{pinto2022sharingiscaring}. 
The selection of the best model for the fine-tuned model is based on its performance on the validation set, using the same process as for pretraining (see Section~\ref{sec:gm_training}). We report the final error as the average over all four seasons.

\subsection{Time Series Foundation Models}
\label{sec:tsfms}

Another approach for modeling thermal dynamics in buildings is Time Series Foundation Models (TSFMs). TSFMs are generally pretrained on large, heterogeneous collections of time series data to learn general-purpose temporal representations that can be used for different downstream tasks. For building thermal dynamics, a TSFM can be applied in a zero-shot setting to directly map input data to a horizon prediction. 
For our experimental comparisons, we selected three frequently used TSFMs.\\

\noindent
\textbf{F1: Chronos.} Chronos-2 \cite{ansari2025chronos2} is a multivariate TSFM that employs a Transformer-based encoder trained mostly on large-scale synthetic data. Chronos-2 enables probabilistic multivariate prediction without requiring a large set of real-world pretraining corpora. The smaller amount of real-world data was supplemented with synthetic data. For our experiments, we used the \textit{amazon/chronos-2} checkpoint.\\

\noindent
\textbf{F2: TimesFM.} TimesFM \cite{das2024timesfm} adopts a patching strategy (i.e., slicing the time series into fixed-size contiguous subseries), but implements a decoder-only autoregressive Transformer that models temporal dynamics by sequentially predicting future patches. TimesFM is explicitly optimized for zero-shot and arbitrary-horizon forecasting, leveraging efficient patch embeddings to scale to long input histories and long-term predictions. Apart from the model described in \cite{das2024timesfm}, we used TimesFM version 2.5, specifically the checkpoint \textit{google/timesfm-2.5-200m-pytorch}, as further described in \cite{timesfm25github}. For our experiments, we used the latest model in the univariate forecasting mode because the multivariate version with XREG is not yet publicly available.\\

\noindent
\textbf{F3: Toto.} Toto \cite{cohen2025toto} is a decoder-only TSFM trained on over 2 trillion time series data points designed for multi-variate zero-shot forecasting tasks. Using a per-variable patch-based causal scaling, Toto claims to be especially efficient towards non-stationary time-series forecasting problems. For our experiments, we use the checkpoint \textit{Toto-Open-Base-1.0} with 151M parameters. \\

We selected a combination of encoder- and decoder-based models. All TSFMs used a combination of real-world and simulated data in their training. For the experiments, all TSFMs are evaluated in zero-shot mode. Usually, encoder-based models only learn the representation of the input and therefore require fine-tuning. Encoder-based models generally learn representations of the input and thus require fine-tuning or a task-specific output layer forecasting head. Nevertheless, we include Chronos as an encoder-based model without fine-tuning, since it performs competitively in this zero-shot setting compared to decoder-based models. All TSFMs had the same input format as the MSGMs, except for Chronos: we added an indexing column for Chronos to the input because an index column was mandatory for the model by design.\\

\subsection{Evaluation}
\label{sec:evaluation}

We use the test set to report errors. The test set is sliced into windows with a lookback of 96 and a forecast horizon of 4, as for the train set. We used the following metrics to evaluate the performance of all models during testing, with $n$ being the number of examples and $h$ being the forecast horizon: 

\begin{equation}
    \text{RMSE} = \sqrt{\frac{1}{n \cdot h} \sum^n_{i=1}\sum^h_{j=1} (T_{in, i+j} - \hat{T}_{in, i+j})^2}
\end{equation}
\begin{equation}
    \text{MAE} = \frac{1}{n \cdot h} \sum^n_{i=1}\sum^h_{j=1} |T_{in, i+j} - \hat{T}_{in, i+j}|
\end{equation}

For our experiments, we used two evaluation types. 
First, we consider a seasonal scenario. There, we evaluate our model's performance on the seasonal test-set, as described in \ref{sec:seasonal_fineuning}. In this context, we assess a fine-tuning and a zero-shot scenario. In both cases, we report the models' performance on the seasonal test sets to ensure comparability. 

Second, we consider an all-in-one scenario. This scenario is applied only to zero-shot performance, as zero-shot requires no further training, and hence, no seasonal overfitting is able to occur.
For the all-in-one scenario, we use the whole time series of the target buildings.

\section{Experiments}
\label{sec:experiments}
This section illustrates the experiments to address our research questions. We first analyze which architectures yield the best performance as an MSGM (section \ref{sec:exp1}) to gain robust insights into the potential of MSGMs. Based on these results, we show to what extent an MSGM can improve forecasting accuracy for thermal dynamics in buildings compared to a single-source TL approach in experiment \ref{sec:exp2}. The experiment in section \ref{sec:exp3} tests how three different TSFMs perform against the MSGMs. The final experiment in Section \ref{sec:exp4} investigates the impact of the number of pretraining sources on MSGM performance and puts the results into context with a TSFM baseline.
In each experiment section, we will first describe the experimental setup and then present the results.
The appendix contains complementary results for each experiment.

\subsection{Architecture Ablation}
\label{sec:exp1}

First, we make an ablation of four general model architectures for MSGMs as described in section \ref{sec:gm_training}. For each architecture, we performed hyperparameter tuning with Optuna. The design and results of the hyperparameter search are detailed in the Appendix. 
For this analysis, we select the all-in-one scenario to gain insights into the influence of the architecture across the whole data set.

\begin{figure}[htbp]
  \centering
 \includegraphics[width=\linewidth]{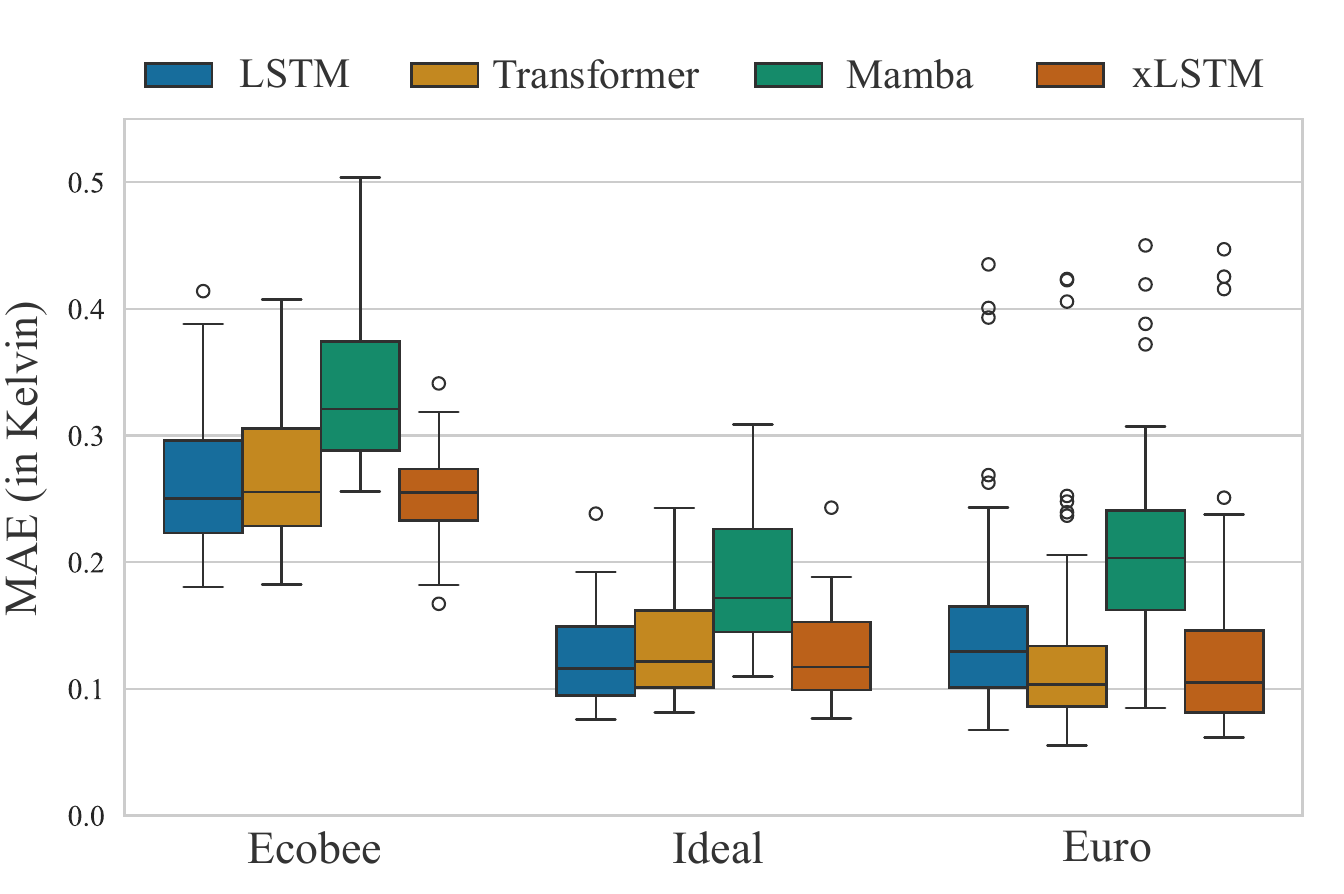}
  \caption{Zero-Shot MAE of all architectures as MSGMs}
  \label{fig:exp01_gm_zero}
\end{figure}

The performance of all architectures across all datasets is shown in the box plots in Figure \ref{fig:exp01_gm_zero}. The results show that 1-hour predictions have an average temperature deviation of approximately 0.1 K for the Ideal and Euro datasets, and approximately 0.25 K for the Ecobee dataset. The boxplots indicate that all architectures perform similarly on average, with Mamba as an outlier. We observe that the choice of the best architecture depends on the dataset: while the Transformer is the best architecture for the European houses dataset, the LSTM is the best choice for the other two datasets. The results also reveal that the choice of deep learning architecture, whether xLSTM, LSTM, or Transformer, plays only a minor role. Further experiments will use the Transformer model for the European Households dataset and the LSTM model for the other two datasets. \\

\subsection{Real-World General Models}
\label{sec:exp2}

In the second experiment, we test how well an MSGM approach can model building thermal dynamics compared to a single-source TL approach on the Ecobee dataset as a real-world representative. For the single-source TL, we randomly choose a source from the Ecobee training set and train an LSTM on it, similar to \cite{GenTL2025, li2024building}. We investigate four fine-tuning scenarios: zero-shot, 0.5 months, 1 month, and 2.5  months. We perform a seasonal fine-tuning as described in section \ref{sec:seasonal_fineuning} on both the single-source model and the MSGM and report the average error across the seasonal test sets for each of the four seasons to ensure robustness of the results.

\begin{figure*}[htbp]
  \centering
  \includegraphics[width=\linewidth]{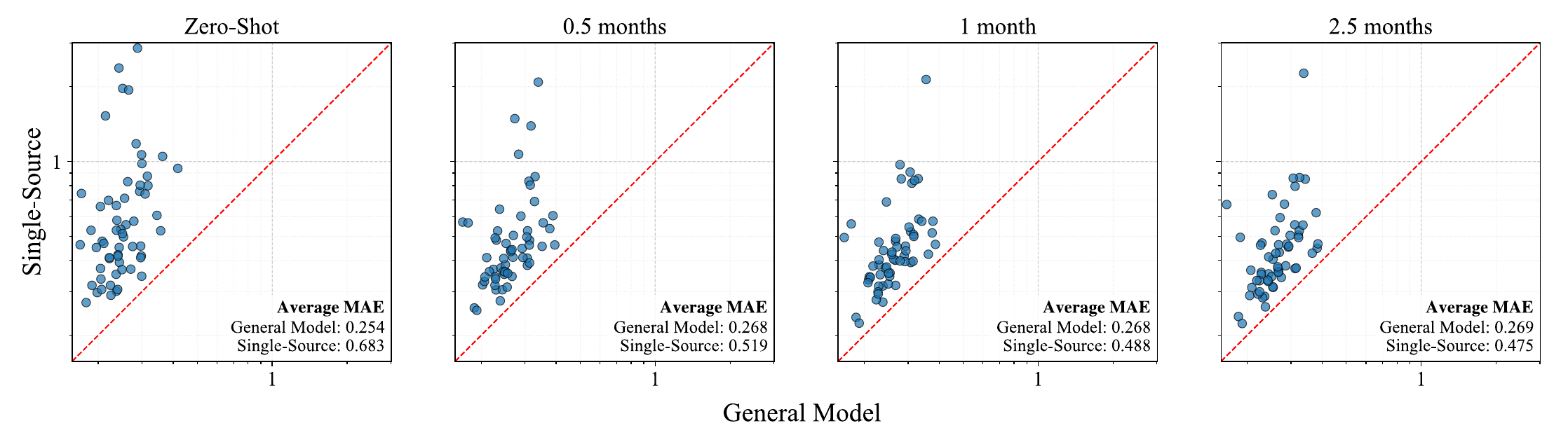}
  \caption{MAE of Single-Source TL versus MSGM for different fine-tunings on real-world data (Ecobee). MSGM beats single-source TL once errors are above the dotted line.}
  \label{fig:single-source-gm}
\end{figure*}

Figure \ref{fig:single-source-gm} illustrates the performance of the single-source TL approach compared to the MSGM approach. Each dot represents one target building from the Ecobee dataset, where the x coordinate represents the MAE of the MSGM and the y coordinate represents the MAE of the single-source TL approach. Both axes are scaled logarithmically for a better representation. A position above the red line indicates that, averaged by season, the MSGM outperforms the single-source approach on the respective building. The results show that the MSGM outperforms the TSFMs across real-world and simulated datasets. Across all four data-sparsity scenarios, the MSGM outperforms the single-source TL on every building, with varying improvements. In addition to that, we find that the overall error of the single-source TL decreases with an increasing number of fine-tuning data. The results also show that on real data, fine-tuning the MSGM decreases its performance regardless of the number of fine-tuning examples. Consequently, we use the MSGMs in zero-shot mode for further experiments.

\subsection{Foundation Models compared to General Models}
\label{sec:exp3}

In our third experiment, we compare three state-of-the-art TSFMs with the MSGM approach on all three datasets. For the MSGM, we chose the best-performing architecture for each dataset as a representative. 

\begin{figure}[htbp]
  \centering
  \includegraphics[width=\linewidth]{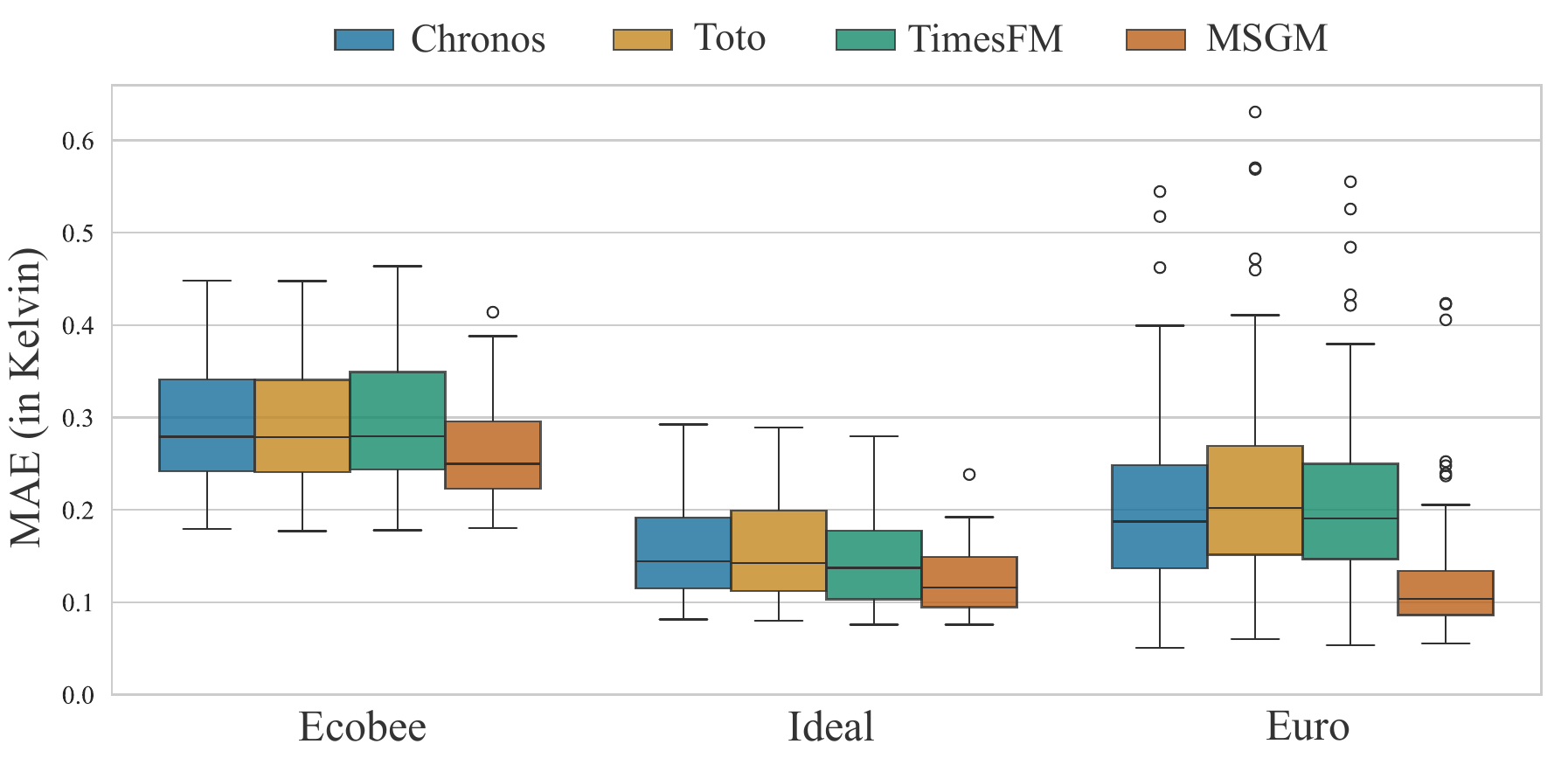}
  \caption{Zero-shot performance: TSFMs vs. the best-performing MSGMs (Transformer for Euro, LSTM for Ecobee and Ideal).}
  \label{fig:exp02_boxplots}
\end{figure}

First, we compare the average performance of the TSFMs to that of the MSGM for each dataset. Figure \ref{fig:exp02_boxplots} illustrates the MAE performance of the MSGM approach with the three TSFMs. The results show that across all datasets, the MSGM outperforms all TSFMs on average. Additionally, TimesFM performs best on the Ideal dataset among the other TSFMs, although it uses only the previous target values as context.

\begin{figure}[htbp]
  \centering
  \includegraphics[width=\linewidth]{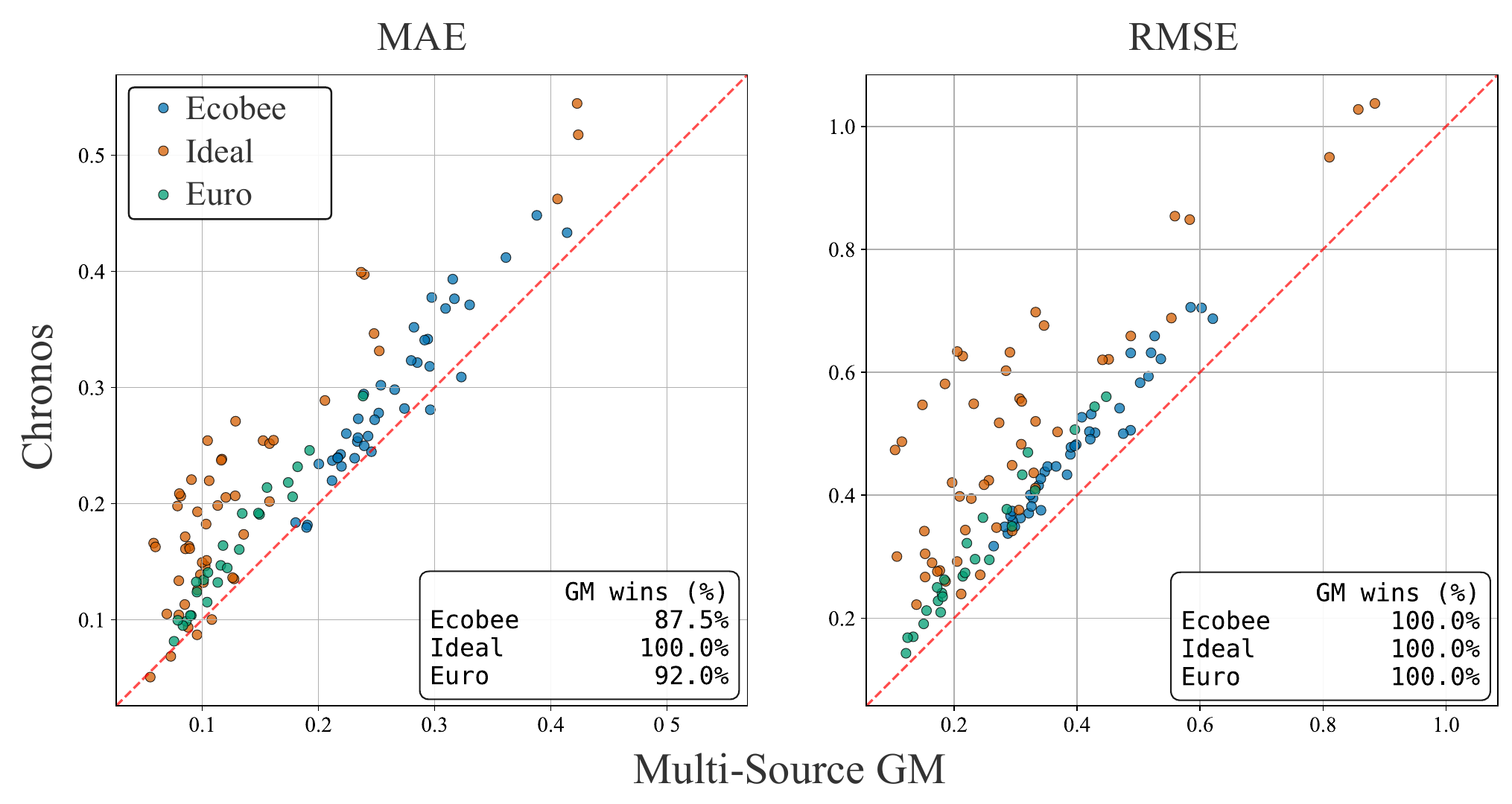}
  \caption{Zero-shot performance: Per-building TSFM compared to MSGM (each data point is a target building)}
  \label{fig:exp02_scatter}
\end{figure}

Second, we make a building-specific comparison of Chronos as a representative of TSFMs with the MSGM approach in Figure \ref{fig:exp02_scatter}. The choice of Chronos was based on having the best average rank across all TSFMs and datasets; however, we observe similar results with the other TSFMs. In the figure, each data point represents one building, with the color indicating the dataset from which it originates. The x-coordinate represents the error for the MSGM approach for the building, and the y-coordinate represents the error of Chronos. Points above the red line indicate that the MSGM achieved better performance on the target building than Chronos. The results show that the MSGM outperforms the TSFM for both metrics and across all datasets. Exceptions for that can be observed for the MAE; we assume that Chronos may perform riskier forecasts, which are not as heavily punished by the MAE as by the RMSE.

\subsection{Same Sensor Building Sparsity}
\label{sec:exp4}

\begin{figure*}[htbp]
  \centering
  \includegraphics[width=\linewidth]{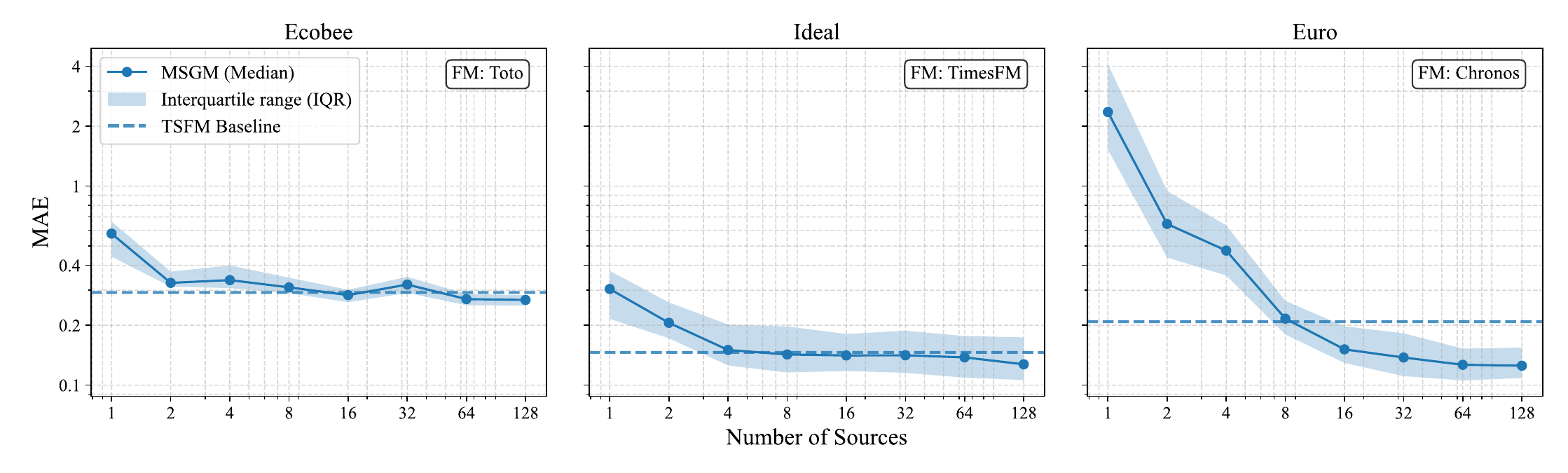}
  \caption{Zero-shot performance: Influence of the number of sources on the MSGM for buildings. The best-performing TSFM with respect to each dataset is used as a baseline.}
  \label{fig:03_gm_size}
\end{figure*}

The fourth experiment investigates how the number of sources used for pretraining a general model influences its overall performance. This experiment directly addresses the same sensor building sparsity problem, a scenario in which only a limited number of buildings equipped with identical sensors may be available. Following the common hypothesis that more sources used during pretraining lead to better prediction performance, leaves open the question on whether there is a threshold concerning the number of sources to which a TSFM can outperform a MSGM. To test this, we train one MSGM for each dataset and each number of sources, ranging from 1 to 128, with power steps of 2. For $n$ sources, we randomly sample $n$ buildings from the source data set and train a general model on them. Afterwards, we evaluate this general model on the target buildings set to obtain an average performance of an MSGM with the respective number of sources. The target set remains the same throughout the experiment. For MSGMs with $\leq$ 16 sources, we repeated the experiment four times and report the best result to ensure robustness and mitigate large performance variance arising from different samples from source buildings.

Figure \ref{fig:03_gm_size} shows the MAE errors of the MSGMs regarding different numbers of sources trained on for all three datasets. The results underline the assumption that more sources generally lead to a better performance for the MSGM, with a decreasing return on investment above 64 sources. In addition, we observe that for around 16 sources in the Ecobee and Euro dataset and 32 sources in the Ideal dataset, the MSGM consistently outperforms the respective TSFM, beating the benchmark on average for the MAE metric. Deviations from that can be seen, for instance, in 32 sources in the Ecobee dataset, which we attribute to an unfavorable random selection. In the Ideal dataset, different rooms within the same building can be selected, resulting in a more homogeneous source dataset. This effect may lead to an increasing number of sources needed to beat the TSFM baseline. For the RMSE metric (see Appendix Figure \ref{fig:appendix_03_number_sources}, we observe that MSGMs outperform TSFMs with between 2 and 8 sources.

\section{Discussion}
\label{sec:discussion}
We present the largest evaluation to date of general models for building thermal dynamics, including both MSGMs and TSFMs. We conduct an ablation study across four state-of-the-art deep learning models to investigate their suitability as an architecture for multi-source general models (MSGM) and compare them with current Time-Series Foundation Models (TSFM). The most effective MSGM architectures were pretrained on a large-scale corpus of data comprising over 500 time series from 300+ buildings, including both simulated and real-world data across diverse climate zones. To ensure practical applicability, we design experiments that address data sparsity and the same sensor building sparsity. Our findings provide the following key insights: \\

\noindent
\textbf{Result 1.} 
Using the Ecobee dataset, we demonstrate that MSGMs outperform traditional single-source Transfer Learning approaches on real-world data. We base our results on four different data sparsity scenarios, ranging from zero-shot to 3 months of data available from the target building. MSGMs reduced averaged forecasting errors by 33\% to 63\%, where scenarios in which less data from the target building are available result in larger improvements. This confirms the conclusion from \cite{GenTL2025}: Given data scarcity, multi-source TL is superior to single-source TL, which can also be applied to real-world buildings. 
Our results also indicate that fine-tuning a dedicated MSGM can actually increase forecasting errors. To our knowledge, such findings have not yet been found in the related literature. We assume that forecasts become less accurate after fine-tuning, either due to overfitting on the season of the validation set.
Therefore, the results suggest deploying MSGMs in zero-shot mode when the available data covers less than three months. \\

\noindent
\textbf{Result 2.} 
Although state-of-the-art TSFMs demonstrate a decent performance, they generally fail to match the accuracy of MSGMs. TSFMs only achieved superior performance in 12.5\% on the Ecobee and 8\% of the buildings in the Euro dataset for the MAE metric. In those cases, we observed that the TSFM outperforms the MSGM only by a small margin. In all other cases, using an MSGM yielded a significant performance increase. Additionally, we also observe that the univariate TimesFM performs comparably and even outperforms the other TSFMs on the Ideal dataset for the MAE metric. This was unexpected, as TimesFM only uses indoor temperature as input, whereas the other models can also take additional important features, such as heat source power, into account. We assume that these results occur because current multivariate TSFMs cannot fully capture the inertia effects of building dynamics and therefore may not utilize additional information as well as an MSGM, for instance.  \\

\noindent
\textbf{Result 3.}
We investigate the impact of the same sensor building sparsity during pretraining on MSGM prediction performance by varying the number of source buildings available for pretraining. A consistent upward trend is observed: the performance of an MSGM increases with the number of buildings used for pretraining. Comparing TSFMs with increasingly better-performing MSGMs, we identify a threshold. If less than 16 same sensor buildings are available or computational resources limit training a MSGM from scratch, we recommend using a TSFM. Otherwise, an MSGM trained on the available source buildings is preferable. This changes when evaluating the performance with the RMSE metric instead of the MAE: in that case, the threshold is that more than 8 source buildings are required for an MSGM to outperform TSFMs. \\

These results align with our research questions. First, to answer our research question 1, we found that MSGMs also outperform single-source TL across a wide range of buildings. Regarding our second question, we observed that MSGMs are superior to the current state-of-the-art TSFMs when modeling building thermal dynamics most of the time. The third research question was answered with result 3: there is a threshold of 16 same-sensor buildings for an MSGM to beat the TSFM benchmark.

\subsection{Computational Efforts}

Another major point in this paper is the computational efficiency of the TSFMs and MSGMs used. Obviously, TSFMs exhibit a larger model corpus, leading to more parameters and, consequently, longer inference times and increased computational effort per forecast. However, this must be viewed from two different perspectives: a TSFM does not require specific pretraining on source building data, which is a major advantage, but it is slower and more computationally costly during inference. 

We conducted our training and inference on one H100 GPU with 94GB VRAM and 80 GB memory. One multi-step forecast for a 1-hour time horizon took around 0.03-0.05s on average for the TSFMs, while the general models required between 0.0002 and 0.0004s (depending on the architecture used). Although the time needed per forecast is around 100 times larger for the TSFM, in the context of forecasts every 15 minutes, this slower inference can likely be neglected.

\subsection{Limitations and Future Work}

This section outlines the limitations of the current study and proposes directions for future research. \\

\noindent
\textbf{Fine-Tuning Strategies for Foundation Models.} While TSFMs were central to our study, they were tested exclusively in a zero-shot mode. Our tests, not included in this paper, indicate that naive fine-tuning on sparse data actually increases error rates, which is an already known challenge in the field \cite{liang2024enablingtsfmcontrastivecurr}. Consequently, three of the four experiments remained in the zero-shot setting, as basic fine-tuning was found to be inferior in Experiment~\ref{sec:exp2}. We made similar observations with the MSGMs approach. To enable building-specific adaptation, future efforts should focus on efficient fine-tuning for TSFMs. This involves implementing advanced fine-tuning techniques such as Low-Rank Adaptation (LoRA) or curriculum learning to stabilize the training process on sparse, domain-specific datasets. \\

\noindent
\textbf{Data Sparsity and Scenario Breadth.} Our experiments regarding single-source Transfer Learning (TL) were restricted to scenarios with a maximum of three months of data. The performance gap between single-source TL and the MSGM approach becomes narrower with more fine-tuning data available. Subsequent research should investigate the gap between single-source TL, MSGMs, and TSFMs over longer time horizons. Furthermore, integrating continual learning frameworks, as already introduced in \cite{raisch2025cl}, would provide a more realistic benchmark to investigate how the trade-off between TSFMs and MSGMs evolves when building parameter shift over time. \\

\noindent
\textbf{Feature-Invariant MSGMs.} TSFMs offer an advantage over MSGMs in that they are less dependent on the specific input features, which motivated their use in our study as an alternative when fewer source buildings are available. However, TSFMs tend to perform poorly compared to MSGMs when a sufficient number of source buildings are accessible. Future work could explore combining the strengths of both approaches by developing feature-invariant MSGMs capable of handling varying numbers and types of input features. Such models could maintain robust performance even for buildings with uncommon or heterogeneous sensor configurations.

\section{Conclusion}
\label{sec:conclusion}

In this paper, we conduct the first comprehensive assessment of general modeling approaches for building thermal dynamics. 
We investigate the performance of three data-efficient approaches for this task, namely single-source Transfer Learning, multi-source general models (MSGMs), and time series foundation models (TSFMs), to determine in what scenario which strategy is best to use. 
We find that the single-source TL strategy yielded the least accurate forecasts for simulated and real-world data. 
Further, we find that \textit{same sensor building sparsity}, meaning only a limited number of buildings with the same sensor are available for pretraining, significantly affects whether to use TSFMs or MSGMs.
Based on our results, we conclude that general models are highly effective for forecasting building thermal dynamics in real-world applications. 
Nevertheless, it is important to note that at least 16 same sensor buildings are required for the effective use of MSGMs; otherwise, TSFMs are the more promising strategy.
Overall, our work sheds light on the when and how of different general modeling approaches and, thus, improves the practical applicability of general models for forecasting building thermal dynamics.



\balance
\bibliographystyle{ACM-Reference-Format}
\bibliography{refs}

\appendix

\section{Appendix}
\label{sec:appendix}
\begin{table*}[h!]
\centering
\caption{Hyperparameter Search Space for the Multi-Source General Models.}
\footnotesize	
\label{tab:z_hyperparam_search_gms}
\begin{tabular}{l 
| ccc 
| ccc 
| ccc 
| ccc}
\toprule

\multirow{2}{*}{\textbf{Hyperparameter}} 
    & \multicolumn{3}{c|}{\textbf{LSTM}} 
    & \multicolumn{3}{c|}{\textbf{Transformer}} 
    & \multicolumn{3}{c|}{\textbf{Mamba}} 
    & \multicolumn{3}{c}{\textbf{xLSTM}} \\
\cmidrule(lr){2-4}
\cmidrule(lr){5-7}
\cmidrule(lr){8-10}
\cmidrule(lr){11-13}
 & \textbf{From} & \textbf{To} & \textbf{Step}
 & \textbf{From} & \textbf{To} & \textbf{Step}
 & \textbf{From} & \textbf{To} & \textbf{Step}
 & \textbf{From} & \textbf{To} & \textbf{Step} \\
\hline

Hidden Size 
 & 64 & 256 & 32
 & 64 & 256 & 32
 & 64 & 256 & 32
 & 64 & 256 & 32 \\

\# Layers
 & 2 & 4 & 1
 & 2 & 4 & 1
 & 2 & 8 & 1
 & 2 & 4 & 1 \\

 Batch Size
 & 32 & 128 & 32
 & 32 & 128 & 32
 & 32 & 128 & 32
 & 32 & 128 & 32 \\

 Learning Rate
 & .0001 & .001 & .0001
 & .0001 & .001 & .0001
 & .0001 & .001 & .0001
 & .0001 & .001 & .0001 \\

 Dropout
 & 0.0 & 0.1 & 0.01
 & 0.0 & 0.1 & 0.01
 & - & - & -
 & 0.0 & 0.1 & 0.01 \\

 \hline

 \# Attention Heads
 & - & - & -
 & 2 & 6 & 1
 & - & - & -
 & 2 & 4 & 1 \\

\bottomrule
\end{tabular}
\end{table*}

\begin{table*}[h!]
\centering
\footnotesize	
\caption{Hyperparameters for all MSGM architectures for each dataset.}
\label{tab:gm_hyperparams}
\begin{tabular}{l | c c c | c c c  | c c c | c c c}
\toprule
\multirow{2}{*}{\textbf{Parameter}} 
    & \multicolumn{3}{c|}{\textbf{LSTM}} 
    & \multicolumn{3}{c|}{\textbf{Transformer}} 
    & \multicolumn{3}{c|}{\textbf{Mamba}} 
    & \multicolumn{3}{c}{\textbf{xLSTM}} \\
    \cmidrule(lr){2-4}
\cmidrule(lr){5-7}
\cmidrule(lr){8-10}
\cmidrule(lr){11-13}
 & \textbf{Ecobee} & \textbf{Ideal} & \textbf{Euro} & \textbf{Ecobee} & \textbf{Ideal} & \textbf{Euro} & \textbf{Ecobee} & \textbf{Ideal} & \textbf{Euro} & \textbf{Ecobee} & \textbf{Ideal} & \textbf{Euro} \\
\hline
Hidden Size & 96   & 128  & 192  & 192 & 128  & 192  & 160   & 128  & 96  & 96 & 128  & 64  \\
\# Layers & 3 & 3  & 3  & 4 & 3  & 3  & 3 & 4 & 3  & 3 & 3  & 3  \\
\hline
Batch Size & 128 & 128  & 96  & 128 & 128  & 128 & 128    & 128  & 96  & 128 & 128 & 128  \\
Use AMP  & True  & True  & True  & True   & True  & True  & True  & True  & True  & False & False  & False  \\
Learning Rate  & .0009 & .001  & .001  & .0001 & .0005  & .0004 & .0006 & .0002  & .0002  & .0002 & .0005 & .0005  \\
Dropout   & 0.0   & 0.0  & 0.0  & 0.0   & 0.04  & 0.04 & -  & - & - & -  & - & - \\
\hline
\# Attention Heads  & - & - & - & 4 & 4  & 4 & - & - & - & 4 & 4 & 2 \\
\hline
Train Time [h] & 4.5 & 1.2 & 6.1 & 9.8 & 2.2 & 8.7 & 14.2 & 6.8 & 18.7 & 17.8 & 4.8 & 20.4 \\
\bottomrule
\end{tabular}
\end{table*}

\begin{figure*}[htbp]
  \centering
  \includegraphics[width=\linewidth]{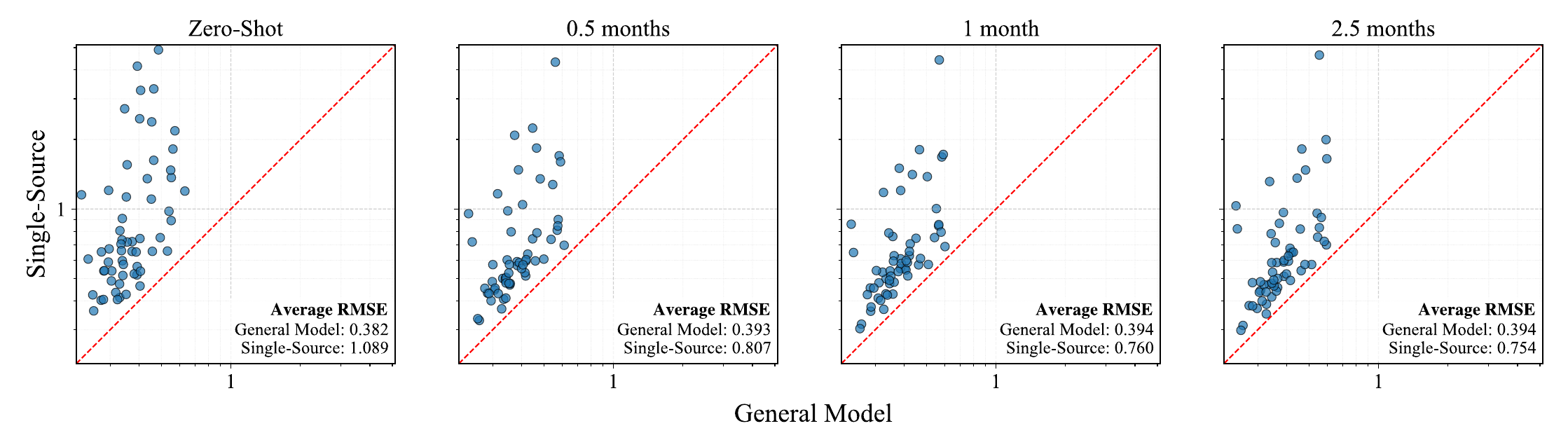}
  \caption{RMSE of Single-Source TL versus MSGM for different fine-tunings on real-world data (Ecobee). MSGM beats single-source TL once errors are above the dotted line.}
  \label{fig:appendix_01_ss}
\end{figure*}

\begin{figure*}[htbp]
  \centering
  \includegraphics[width=\linewidth]{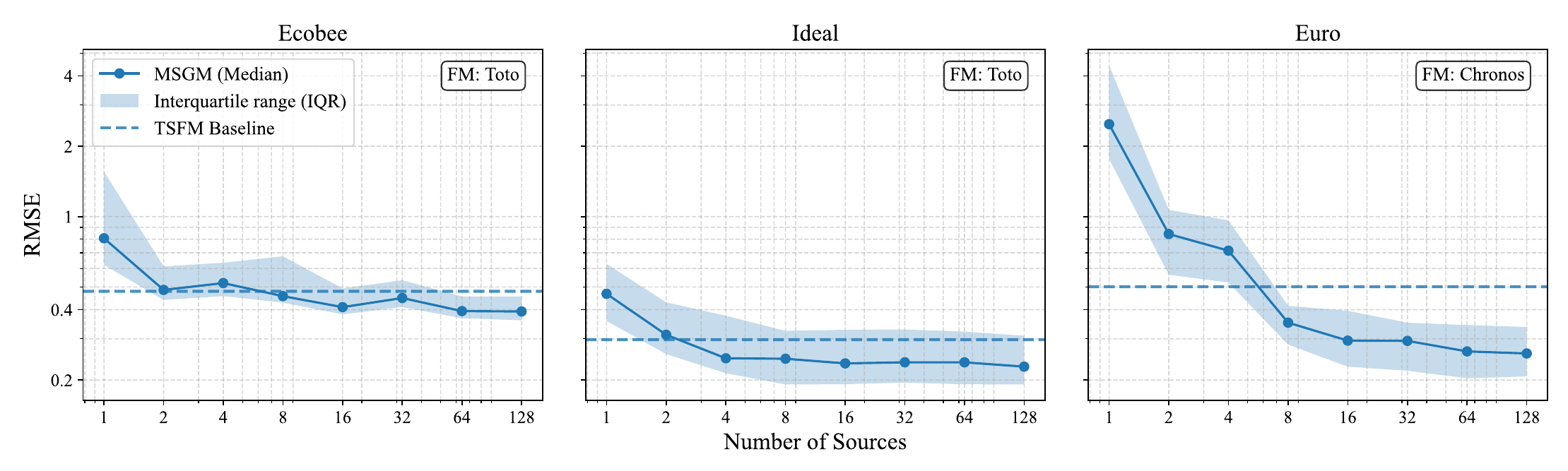}
  \caption{Zero-shot performance RMSE: Influence of the number of sources on the MSGM for buildings. The best-performing TSFM with respect to each dataset is used as a baseline.}
  \label{fig:appendix_03_number_sources}
\end{figure*}

\end{document}